# Optical Control of Nanomechanical Eigenfrequencies and Brownian Motion in Metamaterials

Jinxiang Li[1], Kevin F. MacDonald[1*], and Nikolay I. Zheludev[1, 2]

[1] *Optoelectronics Research Centre, University of Southampton, Highfield, Southampton, SO17 1BJ, UK*

[2] *Centre for Disruptive Photonic Technologies & The Photonics Institute, SPMS, Nanyang Technological University Singapore, 637371, Singapore*

* kfm@orc.soton.ac.uk

Nanomechanical photonic metamaterials provide a wealth of active switching, nonlinear and enhanced light-matter interaction functionalities by coupling optically and mechanically resonant subsystems. Thermal (Brownian) motion of the nanostructural components of such metamaterials leads to fluctuations in optical properties, which may manifest as noise, but which also present opportunity to characterize performance and thereby optimize design at the level of individual nanomechanical elements. We show that Brownian motion in an all-dielectric metamaterial ensemble of silicon-on-silicon-nitride nanowires can be controlled by light at sub-$\mu W/\mu m^2$ intensities. Induced changes in nanowire temperature of just a few Kelvin, dependent upon nanowire dimensions, material composition, and the direction of light propagation, yield proportional changes of several percent in the few-MHz Eigenfrequencies and picometric displacement amplitudes of Brownian motion. The tuning mechanism can provide active control of frequency response in photonic metadevices and may serve as a basis for bolometric, mass and micro/nanostructural stress sensing.

**Introduction**

By virtue of their low mass and fast (MHz-GHz frequency) response times, nanomechanical oscillators actuated and/or interrogated by light are of fundamental and applied interest in numerous applications, ranging from mass and force sensors to photonic data processing and quantum ground state measurements[1-12]. As the dimensions of such systems decrease their thermal (i.e. Brownian) motion assumes increasing importance. By adding noise to induced/controlled movements that underpin the functionality it can constrain performance, but it also presents opportunity by directly linking observable (far-field optical) properties to geometry, composition and temperature at the nanoscale. We show here that such motion can be optically controlled at $\mu W/\mu m^2$ intensities in nanomechanical photonic metamaterials. In an array of mechanically independent and (by design) alternately dissimilar dielectric nanowires, which are at the same time of identical bilayer (i.e. asymmetric) material composition and part of an optically resonant ensemble subject to the fundamental constraint of linear transmission reciprocity, dependences of motion Eigenfrequencies and picometric displacement amplitudes on local light-induced temperature changes can be accurately determined.

**Results and Discussion**

In the present study, we employ an all-dielectric metamaterial comprising pairs of dissimilar (by length and width) silicon nano-bricks on a free-standing array of flexible silicon nitride nanowires (Fig. 1a). It is fabricated on a 200 nm thick $Si_3N_4$ membrane coated by plasma enhanced chemical vapor deposition with a 115 nm thick layer of amorphous Si. This bilayer is then structured by focused ion beam milling to define rows of alternately short, narrow (720 nm × 210 nm) and long, wide (780 nm × 300 nm) nano-bricks in the Si layer on parallel 21.1 µm long nanowire beams cut through the $Si_3N_4$ layer, with a gap size between neighboring nanowires of 170 nm.

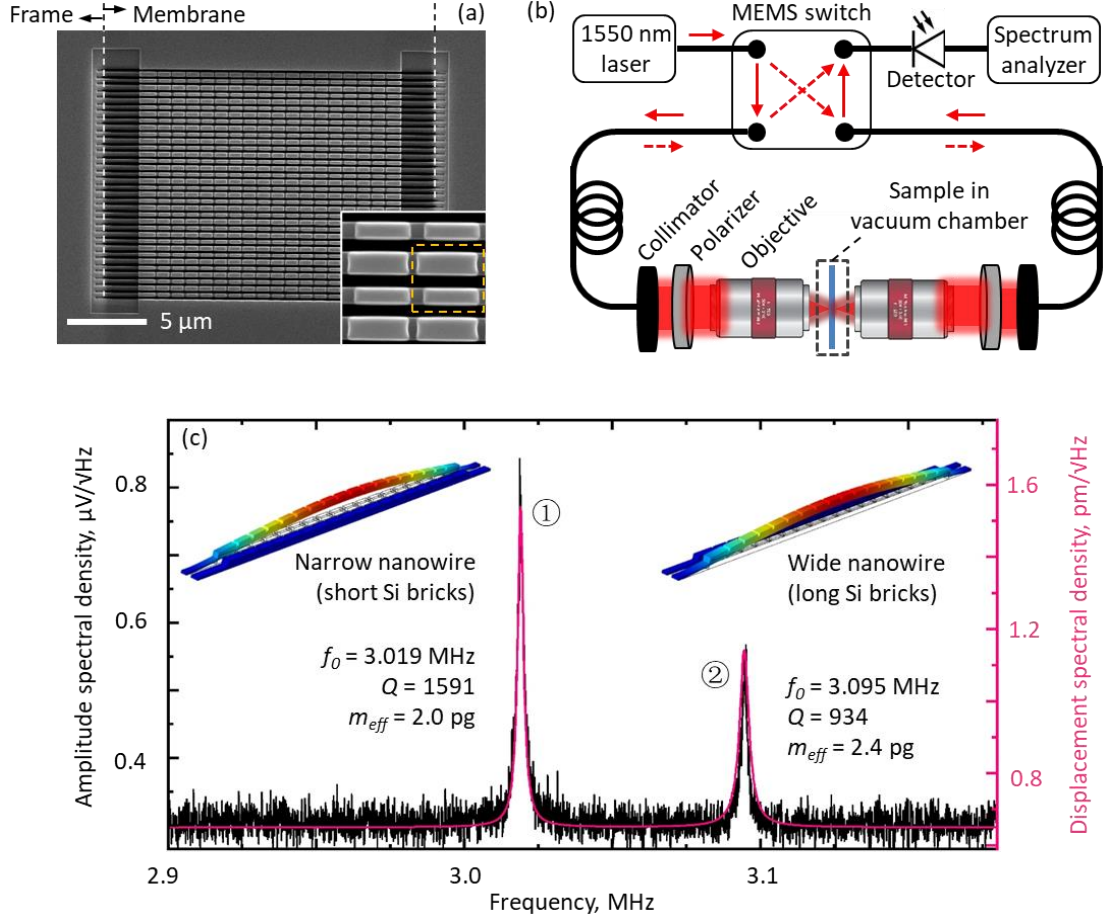

**Figure 1. Detecting thermal (Brownian) motion of nanowires within an all-dielectric nanomechanical metamaterial.** (a) Scanning electron microscope image of the metamaterial, fabricated on a 20.5 μm wide free-standing silicon nitride membrane. The inset enlarged section shows detail of the supported silicon nano-bricks – the dashed line denotes a unit cell of the structure. (b) Schematic of experimental apparatus for recording frequency spectra of metamaterial transmission. Other than between the two collimators, light is carried in polarization-maintaining single-mode optical fiber, with the MEMS switch providing for inversion of the light propagation direction through the sample. (b) Exemplar measurement of optical transmission amplitude spectral density [for light incident on the silicon nitride side of the sample at a power level of 15.9 μW], showing a pair of peaks associated with the mechanical resonances of two individual nanowires within the array: ①/② a narrower/wider wire decorated with shorter/longer Si bricks. The overlaid magenta curve and calibrated displacement spectral density scale [to the right-hand side] are obtained by fitting Eq. (1) to the experimental data. Derived values of $f_0$, $Q$ and $m_{eff}$ are shown inset.

The metamaterial structure supports a near-infrared closed mode optical resonance[13] at a wavelength of 1542 nm (see Supplementary Fig. S1), underpinned by the excitation of antiparallel displacement currents in adjacent dissimilar silicon nano-bricks by incident light polarized parallel to the long axis of the bricks. In the vicinity of this optical resonance, thermal (Brownian) motion of the nanowires – mutual positional fluctuations of pico- to nanometric amplitude – translate to fluctuations of metamaterial transmission (of order 0.1%) at their few-MHz natural mechanical resonance frequencies[14]. These thermomechanical oscillations are detected as peaks in frequency spectra of transmission amplitude spectral density (Fig 1b, c): the metamaterial is mounted in a vacuum chamber at a pressure of $4\times10^{-3}$ mbar (to exclude air damping of mechanical motion). This is located between a confocal pair of 20× (NA 0.4) microscope objectives, via which incident light at a wavelength of 1550 nm is focused onto the sample (to a spot of diameter ~5 µm) and transmitted light is collected. The arrangement includes a fiber-optic MEMS switch to enable transmission measurements in both directions through the sample without disturbance of its position/alignment relative to the beam path. (The 'forward' direction of light propagation is designated as that for which light is incident on the Si side of the sample; 'backward' the $Si_3N_4$ side.)

Figure 1c shows a representative measurement of optical transmission amplitude spectral density (ASD), in which peaks associated with the fundamental out-of-plane flexural modes of a pair of individual nanowires – one narrow and one wide, decorated respectively with short and long Si nano-bricks – are seen. (Attribution to this oscillatory mode is confirmed through computational modelling – see Supplementary Information).

Nanowire displacement ASD can be expressed as[15, 16]:

$$\sqrt{S(f)} = \sqrt{\frac{k_B T f_0}{2\pi^3 m_{eff} Q[(f_0^2 - f^2)^2 + (ff_0/Q)^2]}} \quad (1)$$

where $k_B$ is the Boltzmann constant, $T$ is temperature and, for each mode, $m_{eff}$, $f_0$ and $Q$ are respectively the effective mass, natural frequency and quality factor of the oscillator. Experimental data can thus be calibrated – the vertical scale in Fig. 1c converted from signal measured in µV to nanowire displacement in picometres, by fitting Eq. (1) to the data. Specifically, we fit a linear superposition of two instances of the expression, one for each of the spectral peaks with co-optimized values of $f_0$, $Q$ and $m_{eff}$. Root mean square (RMS) thermal motion amplitudes can then be evaluated as the square root of an integral of power spectral density ($ASD^2$) over frequency. These calculations yield amplitudes of 76 and 67 pm respectively for the lower and higher frequency peaks in Fig. 1c, which compare extremely well with analytical values of 76 and 68 pm derived from energy equipartition theorem[17]:

$$\langle z \rangle = \sqrt{\frac{\kappa_B T}{4\pi^2 m_{eff} f_0^2}} \quad (2)$$

Figure 2 shows how the Brownian motions characteristics of nanowires, as manifested in the ASD of optical transmission, depend upon (i.e. can be controlled by tuning) incident laser power, and how this dependence differs for the two directions of incident light propagation. With increasing laser power, mechanical Eigenfrequencies red-shift (Fig. 3a) and RMS displacement amplitudes increase (Fig. 3b), both in direct proportion and more rapidly for the backward propagation direction. The behaviors are consistent with a photothermal tuning mechanism, whereby laser-induced heating decreases tensile stress in the nanowires. The effect is more pronounced for the backward direction of light propagation because, while forward and backward transmission levels are identical (as they must be in a linear, reciprocal medium), reflectivity and absorption are not (Fig. S1).

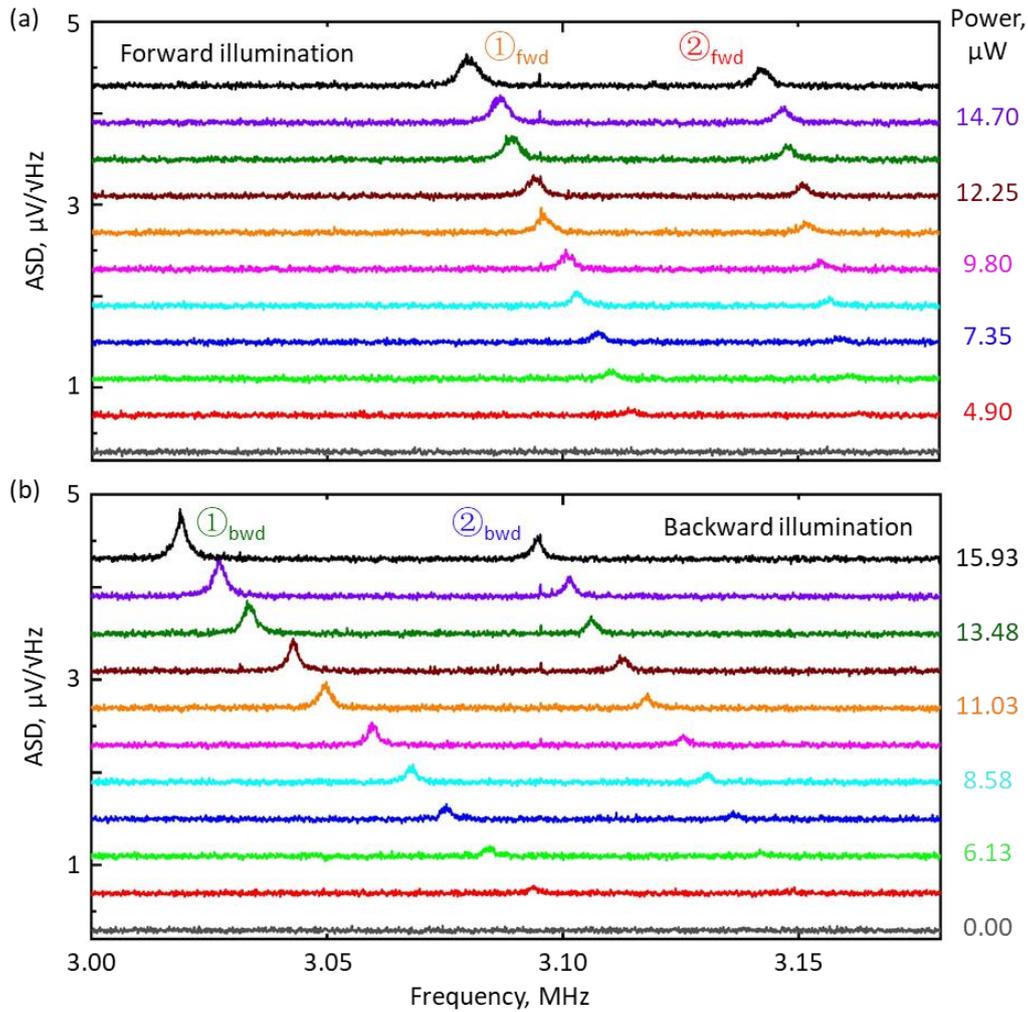

**Figure 2. Optical control of nanowire Eigenfrequencies.** Transmission amplitude spectral density, showing peaks ① and ② as assigned in Fig. 1c, for opposing directions of light propagation through the sample – (a) forward and (b) backward [light incident respectively on the silicon and the silicon nitride side], and for a range of laser power levels [as labelled].

An analytical model for optical control of thermomechanical (Brownian) motion resonances, tightly constrained by the requirement to describe the properties of two independent, similar (related) but not identical oscillators – narrow and wide $Si_3N_4/Si$ bilayer nanowires, under two similar (related) but not identical regimes of optical excitation – forward and backward directions of illumination, provides for accurate quantitative evaluation of light-induced temperature changes in the individual nanowires and corresponding relationships between their resonance frequencies/amplitudes, illumination conditions and the optical properties of the metamaterial array. From Euler-Bernoulli beam theory[18], the stress-dependent fundamental frequency of a doubly clamped beam of homogenous rectangular cross-section is:

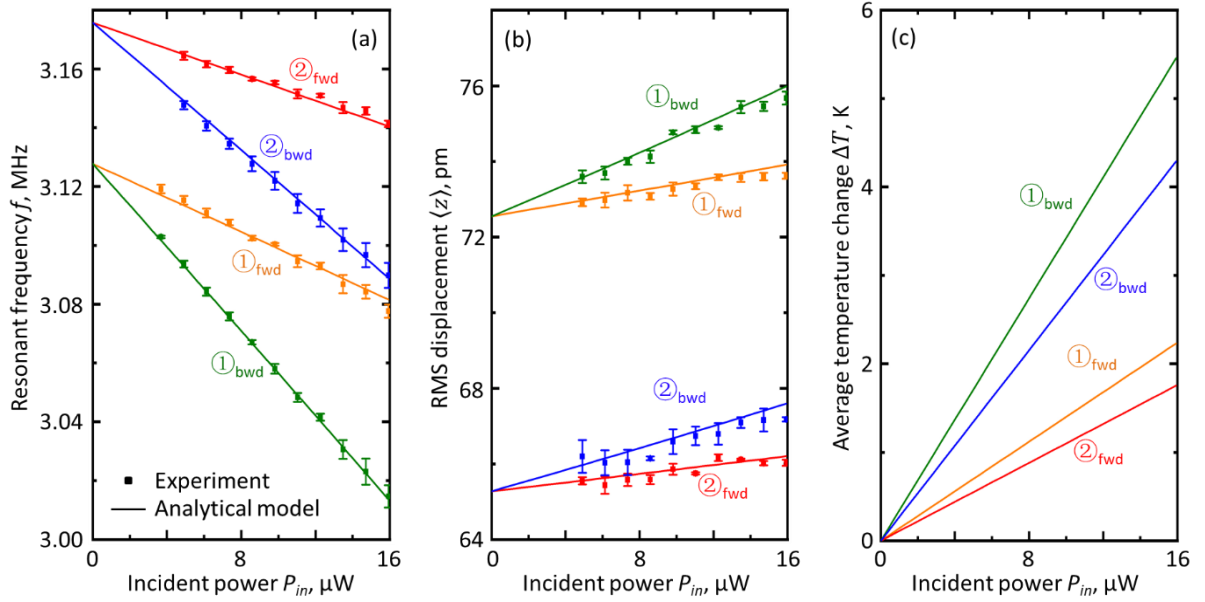

**Figure 3. Optical control of nanowire Eigenfrequencies and Brownian motion amplitudes.** Dependences for peaks ① and ② in Fig. 2 [i.e. for narrow and wide nanowires as identified in Fig. 1c, under nominally forward and backward directions of illumination] of (a) resonance frequency, (b) RMS displacement amplitude, and (c) light-induced nanowire temperature change on incident laser power [total power incident on the metamaterial sample]. Square symbols are experimental data points, with error bars given by the standard deviation over three repeated measurement cycles. Solid lines are derived from an analytical description of the photothermal tuning mechanism via a simultaneous best-fit to the four experimental datasets in (a).

$$f_0 = 1.03 \frac{t}{L^2} \sqrt{\frac{E}{\rho}\left(1 + \frac{\sigma L^2}{3.4E}\right)} \qquad (3)$$

where $t$ and $L$ are the thickness and length of the beam, $E$ is Youngs's modulus, $\rho$ is density and $\sigma$ is tensile stress along the beam length. The temperature-dependence of stress can be expressed in the form[19]:

$$\sigma = \sigma_0 - \alpha E \Delta T \qquad (4)$$

where $\sigma_0$ is the stress at ambient temperature, $\Delta T$ is the difference between average beam and ambient temperatures, and $\alpha$ is the beam's thermal expansion coefficient.

Assuming the presence of a heat source uniformly distributed over the rectangular cross-section at the mid-point of the beam, while the two ends are held at ambient temperature ($T_0 = 298$ K), the equilibrium difference $\Delta T$ between average beam and ambient temperatures is[19]:

$$\Delta T = \frac{PL}{8\kappa A} \qquad (5)$$

Where $\kappa$ is thermal conductivity, $A$ is the cross-sectional area, and $P$ is the power of the heat source.

For the purpose of applying these analytical expressions to the present case, we approximate the silicon nitride nanowires decorated with silicon nano-bricks as simple rectangular-section beams of the same length with effective values of $E$, $\rho$, $\alpha$, $\kappa$, and $A$ derived from the material parameters of Si and $Si_3N_4$ and the volume fractions of the two materials in each nanowire type (see Supplementary Information). $P$ is taken as the optical power absorbed by a nanowire: $P_{abs} = \mu\gamma P_{in}$, where $P_{in}$ is the total power incident on the metamaterial, $\gamma$ is the absorption coefficient of the metamaterial, and $\mu$ is the 'absorption cross-section' of an individual nanowire.

By fitting equations (3)-(5) simultaneously to all four sets of experimental datapoints in Fig. 3a, for the dependences of Brownian motion Eigenfrequency on incident laser power, under constraints that:

(i) $\mu$ must be identical for forward and backward directions of illumination for a given nanowire (i.e. the same nanowire will intercept the same fraction of incident light in both directions);

(ii) $\gamma$ must be identical for the two nanowires for a given illumination directions (i.e. as a metamaterial ensemble property, absorption can only have a single value in each direction);

(iii) $\sigma_0$ must take a single fixed value (ensuring degeneracy of nominally forward- and backward-illumination zero-power resonant frequencies for each type of nanowire);

we obtain the four solid lines plotted in each panel of Fig. 3. The Eigenfrequency fitting (Fig. 3a) is extremely good and yields zero-power (i.e. ambient temperature) resonant frequencies of 3.18 3.13 MHz for the wider and narrower nanowires respectively, giving a value for intrinsic tensile stress of $\sigma_0 = 5.6$ MPa. Derived values of $\mu$ – 4.9% and 5.6% respectively for the narrower and wider nanowires – are consistent, to a first approximation, with their areas of geometric intersection with the ~5 µm diameter incident laser spot. However, they are not in proportion simply to the ratio of nanowire widths (0.7:1). This is because the near-field distribution of electromagnetic field around the metamaterial at resonance is not homogenous, and the absorption cross section of constituent nanowires is therefore not expected to be directly proportional to its geometric cross-section. Derived values of $\gamma$ – 10.6% and 26.0% respectively for the forward and backward directions of light propagation – correspond very well to measured (far-field) values of metamaterial absorption at 1550 nm (Fig. S1a): 13.5% and 23.7%.

Light-induced changes in nanowire temperature (Fig. 3c) depend upon the direction of illumination and nanowire dimensions – which is to say, upon the strength of optical absorption and the rate at which heat is dissipated (the latter being lower for the nanowire of smaller cross-section). The narrow nanowire changes temperature at a rate of 27 K/µW of absorbed power, and the wide nanowire at 19 K/µW. These derived dependences of induced temperature change $\Delta T$ on laser power map to the theoretical dependences of RMS Brownian motion amplitude presented in Fig. 3b via Eq. (2), using values of $m_{eff}$ for the two nanowires established in the above (Fig. 1c) calibration of displacement spectral density. The picometrically accurate correlation with experimental data points separately derived from integrals of PSD over frequency is remarkably good.

**Conclusion**

In summary, we have shown that fluctuations in the resonant optical properties of a photonic metamaterial, which are associated with the mechanically resonant Brownian motion of its constituent elements, can be controlled by light at sub-µW/µm² intensities. In an all-dielectric metamaterial ensemble of free-standing silicon nitride nanowires (mechanical oscillators) supporting an array of silicon nano-bricks (optical resonators), the few-MHz Eigenfrequencies and picometric amplitudes of individual nanowires' motion are directly proportional to incident laser power – changing respectively, in consequence of light-induced heating, by up to 0.7% and 0.9% per K.

An analytical model for the photothermal tuning mechanism, simply but strictly constrained by the requirements of optical transmission reciprocity in a linear medium, links the local, nanoscopic properties and behaviors of individual nanowires (i.e. at sub-wavelength scale) to the far-field optical properties of the (micro/macroscopic) metamaterial ensemble. It provides for accurate evaluation of light-induced changes in nanowire temperature and of ambient condition (zero-illumination) tensile stress and Brownian motion Eigenfrequencies and displacement amplitudes.

The ability to finely tune the nanomechanical resonance characteristics of photonic metamaterials may be beneficial in a variety of metadevice applications where, for example, the frequency of nanostructural oscillation is required to match (or avoid matching) another frequency, such as that of a pulsed laser. The fact that tuning characteristics can (as here) depend strongly upon the direction of light propagation through a metamaterial by simple virtue of bi-layer material composition (leading to different levels of reflection and absorption for light incident on opposing sides) may find application in devices to favor/select a single direction of propagation. The accurately quantifiable sensitivity of optical response to nanomechanical properties in such structures also suggests applications to bolometric sensing and detection of changes in mass (e.g. through adsorption/desorption) or micro/nanostructural stress.

**Data availability:** Following a period of embargo, the data from this paper will be available from the University of Southampton ePrints research repository.

**Acknowledgments:** This work was supported by the Engineering and Physical Sciences Research Council (EPSRC), UK (grant numbers EP/M009122/1 and EP/T02643X/1); the Ministry of Education, Singapore (MOE2016-T3-1-006) and the China Scholarship Council (201708440254). The authors would also like to thank to Jun-Yu Ou for contributions to the design of experimental apparatus.

**References**

1.  J. Chaste, et al., *"A nanomechanical mass sensor with yoctogram resolution,"* Nature Nanotechnology **7** (5), 301-304 (2012).

2.  D. Rugar, et al., *"Single spin detection by magnetic resonance force microscopy,"* Nature **430** (6997), 329-332 (2004).

3.  P. X.-L. Feng, et al., in *Springer Handbook of Nanotechnology* (Springer, 2017), pp. 395-429.

4.  D. Van Thourhout and J. Roels, *"Optomechanical device actuation through the optical gradient force,"* Nat. Photon. **4**, 211-217 (2010).

5.  B. J. Eggleton, et al., *"Brillouin integrated photonics,"* Nature Photonics **13** (10), 664-677 (2019).

6. N. I. Zheludev and E. Plum, *"Reconfigurable nanomechanical photonic metamaterials,"* Nat. Nanotech. **11**, 16-22 (2016).

7. A. Karvounis, et al., *"Giant electro-optical effect through electrostriction in a nano-mechanical metamaterial,"* Adv. Mater. **31**, 1804801 (2019).

8. Q. Zhang, et al., *"Electrogyration in metamaterials: Chirality and polarization rotatory power that depend on applied electric field,"* Adv. Opt. Mater., 2001826 (2020).

9. F. Ye, et al., *"Atomic layer $MoS_2$ graphene van der Waals heterostructure nanomechanical resonators,"* Nanoscale **9** (46), 18208-18215 (2017).

10. A. W. Rodriguez, et al., *"The Casimir effect in microstructured geometries,"* Nat. Photon. **5**, 211-221 (2011).

11. K. E. Khosla, et al., *"Displacemon electromechanics: how to detect quantum interference in a nanomechanical resonator,"* Phys. Rev. X **8** (2), 021052 (2018).

12. A. D. O'Connell, et al., *"Quantum ground state and single-phonon control of a mechanical resonator,"* Nature **464** (7289), 697-703 (2010).

13. V. A. Fedotov, et al., *"Sharp trapped-mode resonances in planar metamaterials with a broken structural symmetry,"* Phys. Rev. Lett. **99**, 147401 (2007).

14. J. Li, et al., *"Thermal fluctuations of the optical properties of nanomechanical photonic metamaterials,"* Adv. Opt. Mater. (in press).

15. B. D. Hauer, et al., *"A general procedure for thermomechanical calibration of nano/micro-mechanical resonators,"* Ann. Phys. - New York **339**, 181-207 (2013).

16. A. A. Clerk, et al., *"Introduction to quantum noise, measurement, and amplification,"* Reviews of Modern Physics **82** (2), 1155-1208 (2010).

17. A. N. Cleland, *Foundations of nanomechanics: from solid-state theory to device applications*. (Springer Science & Business Media, 2013).

18. A. Bokaian, *"Natural frequencies of beams under tensile axial loads,"* Journal of Sound and Vibration **142** (3), 481-498 (1990).

19. S. Yamada, et al., *"Photothermal Infrared Spectroscopy of Airborne Samples with Mechanical String Resonators,"* Analytical Chemistry **85** (21), 10531-10535 (2013).

# Optical Control of Nanomechanical Eigenfrequencies and Brownian Motion in Metamaterials: Supplementary Information

**Metamaterial optical properties**

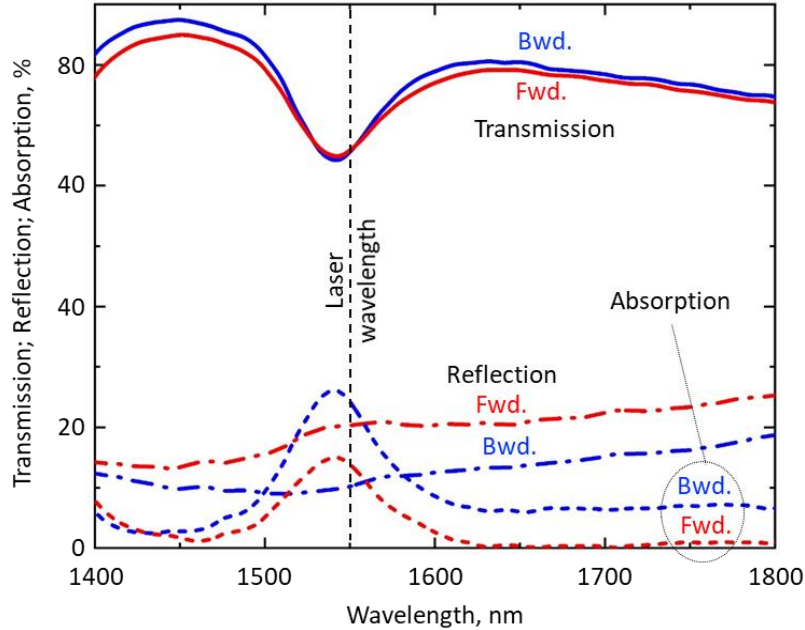

**Fig. S1:** Spectral dispersion of metamaterial transmission $T$ and reflection $R$ measured using a microspectrophotometer, and absorption $A$ calculated as [100-($T$+$R$)], for incident light polarized parallel to the nanowires. Spectra are measured for the two different directions of light propagation (nominally forward and backward) through the sample by removing it from the instrument, turning it over, and reinserting it. Resulting positional alignment imperfections account for the small discrepancies between the two transmission spectra (which would be identical in an ideally reciprocal pair of measurements).

**Nanowire geometry and mechanical resonance frequencies**

The metamaterial is fabricated on a 185 nm thick, $Si_3N_4$ membrane coated with 100 nm of amorphous Si, structured to define rows of alternately narrow (type ①) and wide (type ②) nanowires each supporting a row of eighteen short and long, respectively, Si nano-bricks, according to the dimensional schematic in Fig. S2.

Nanowire mechanical properties are simulated using the structural mechanics module in COMSOL Multiphysics (finite element method). For the ideal rectilinear geometry of Fig. S2, and densities and Young's moduli of silicon nitride and silicon as in Table S1 below, Eigenfrequencies of the fundamental out-of-plane flexural modes are estimated as 3.04 and 3.13 MHz respectively for the type ① (narrower) and type ② (wider) nanowires.

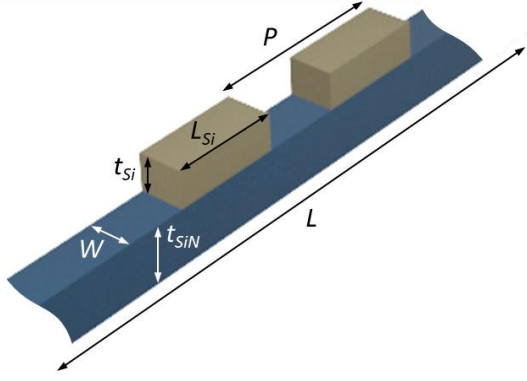

| | Nanowire type ① | Nanowire type ② |
|---|---|---|
| Nanowire length $L$, µm | 21.0 | |
| Width $W$, nm | 210 | 300 |
| $Si_3N_4$ thickness $t_{SiN}$, nm | 200 | |
| Si thickness $t_{Si}$, nm | 115 | |
| Si brick length $L_{Si}$, nm | 720 | 780 |
| Si brick period $p$, nm | 910 | |

**Fig. S2:** Dimensional schematic of a section of an all-dielectric nanomechanical metamaterial nanowire element

These are sufficiently close to each other and to the measured frequencies (Fig. 1c) as to confirm that the Brownian motion peaks observed in experiment are associated with this mode of oscillation – discrepancies being accounted for by manufacturing imperfections and internal stress within the silicon nitride. (For comparison, Eigenfrequencies of the fundamental in-plane flexural modes are much higher, at around 4.5 MHz.)

**Table S1:** Material properties

| | Si | $Si_3N_4$ |
|---|---|---|
| **Density $\rho$, kg.m$^{-3}$** | 2330 | 3100 |
| **Young's Modulus $E$, GPa** | 165 | 260 |
| **Thermal expansion coefficient $\alpha$, K$^{-1}$** | $1.0 \times 10^{-6}$ | $2.8 \times 10^{-6}$ |
| **Thermal conductivity $\kappa$, Wm$^{-1}$K$^{-1}$** | 1.5 | 2.0 |

**Nanowire effective medium parameters**

For the purpose of analytically modeling the photothermal tuning of Brownian motion resonances, the nanowires are assumed to be of simple rectangular cross-section and homogenous material composition. The real value is taken for length $L$ and physical properties are assigned effective values (Table S2) based upon the volume fractions of Si and $Si_3N_4$ present, according to the expression:

$$X_e = \frac{X_{Si}V_{Si} + X_{SiN}V_{SiN}}{V_{Si} + V_{SiN}}$$

where $X$ is density $\rho$, Young's modulus $E$, the thermal expansion coefficient $\alpha$, or thermal conductivity $\kappa$ (values of $X_{SiN}$ and $X_{Si}$ being given in Table S1). $V_{SiN}$ and $V_{Si}$ are respectively the volumes of silicon and silicon nitride, evaluated from electron microscopic measurements to encompass real-sample deviations from the ideal geometry of Fig. S2: specifically, the fact that the Si nano-bricks at each end of the row on each nanowire are of slightly different length, and that the sections of bare silicon nitride at each end, between the nano-brick array and the anchor points, are over-milled to a reduced thickness of 80 nm. Effective cross-sectional area $A_e$ is then evaluated as $(V_{SiN} + V_{Si})/L$. Effective thickness $t_e$ is employed as a fitting parameter, taking derived best-fit values of 139 and 142 nm respectively for type ① and ② nanowires,.

Table S2: Nanowire effective medium parameters

|  | Nanowire type ① | Nanowire type ② |
|---|---|---|
| **Density $\rho_e$, kg.m$^{-3}$** | 2870 | 2859 |
| **Young's Modulus $E_e$, GPa** | 232 | 230 |
| **Thermal expansion coefficient $\alpha_e$, K$^{-1}$** | 2.263×10$^{-6}$ | 2.237×10$^{-6}$ |
| **Thermal conductivity $\kappa_e$, Wm$^{-1}$K$^{-1}$** | 1.851 | 1.844 |
| **Cross-sectional area $A_e$, nm$^2$** | 52760 | 76945 |